\documentclass[twocolumn,trackchanges]{aastex61}
\newcommand{\PIMA}{$\cal P\hspace{-0.067em}I\hspace{-0.067em}M\hspace{-0.067em}A\hspace{-0.1em}$ }
\newcommand{\ntab}[2]{ \multicolumn{1}{#1}{#2} }
\newcommand{\nntab}[2]{ \multicolumn{2}{#1}{#2} }
\newcommand{\nnntab}[2]{ \multicolumn{3}{#1}{#2} }

\newcommand{\nnnnntab}[2]{ \multicolumn{5}{#1}{#2} }

\newcommand{\ns}{\normalsize}
\usepackage{xcolor}
\definecolor{Dred}{rgb}{0.312,0.070,0.070}
\definecolor{Dblue}{rgb}{0.070,0.070,0.312}

\newcounter{note}
\let\oldmarginpar\marginpar
\renewcommand\marginpar[1]{\-\oldmarginpar[\raggedleft\footnotesize #1]%
{\raggedright\footnotesize #1}}

\shorttitle{VLBI ecliptic plane survey: VEPS--1}
\shortauthors{Shu et al.}

\begin{document}

\title{VLBI ecliptic plane survey: VEPS--1}

\correspondingauthor{Fengchun Shu}
\email{sfc@shao.ac.cn}

\author{Fengchun Shu}
\affil{Shanghai Astronomical Observatory, Chinese Academy of Sciences, Shanghai 200030, China}
\affil{Key Laboratory of Radio Astronomy, Chinese Academy of Sciences, China}

\author{Leonid Petrov}
\affiliation{Astrogeo Center, Falls Church, VA 22043, USA}

\author{Wu Jiang}
\affil{Shanghai Astronomical Observatory, Chinese Academy of Sciences, Shanghai 200030, China}

\author{Bo Xia}
\affil{Shanghai Astronomical Observatory, Chinese Academy of Sciences, Shanghai 200030, China}
\affil{Key Laboratory of Radio Astronomy, Chinese Academy of Sciences, China}

\author{Tianyu Jiang}
\affil{Shanghai Astronomical Observatory, Chinese Academy of Sciences, Shanghai 200030, China}

\author{Yuzhu Cui}
\affil{Shanghai Astronomical Observatory, Chinese Academy of Sciences, Shanghai 200030, China}

\author{Kazuhiro Takefuji}
\affiliation{Kashima Space Technology Center, National Institute of Information and Communications Technology, Kashima, Japan}

\author{Jamie McCallum}
\affiliation{University of Tasmania, Private Bag 37, 7001 Hobart, Australia}

\author{Jim Lovell}
\affiliation{University of Tasmania, Private Bag 37, 7001 Hobart, Australia}

\author{Sang-oh Yi}
\affiliation{National Geographic Information Institute, Suwon-si, Gyeonggi-do 16517, Korea}

\author{Longfei Hao}
\affil{Yunnan Astronomical Observatory, Chinese Academy of Sciences, Kunming 650216, China}

\author{Wenjun Yang}
\affil{Xinjiang Astronomical Observatory, Chinese Academy of Sciences, Urumqi 830011, China}

\author{Hua Zhang}
\affil{Xinjiang Astronomical Observatory, Chinese Academy of Sciences, Urumqi 830011, China}

\author{Zhong Chen}
\affil{Shanghai Astronomical Observatory, Chinese Academy of Sciences, Shanghai 200030, China}
\affil{Key Laboratory of Radio Astronomy, Chinese Academy of Sciences, China}

\author{Jinling Li}
\affil{Shanghai Astronomical Observatory, Chinese Academy of Sciences, Shanghai 200030, China}

\begin{abstract}

  We present here the results of the first part of the VLBI Ecliptic Plane
Survey (VEPS) program. The goal of the program is to find all compact sources within
$7.5^\circ$ of the ecliptic plane which are suitable as calibrators for
anticipated phase referencing observations of spacecraft and determine their
positions with accuracy at the 1.5~nrad level. We run the program in two
modes: the search mode and the refining mode. In the search mode, a complete
sample of all sources brighter than 50 mJy at 5 GHz listed in the
Parkes-MIT-NRAO (PMN) and Green Bank 6~cm (GB6) catalogs, except those
previously detected with VLBI, is observed. In the refining mode, the positions of
all ecliptic plane sources, including those found in the search mode, are
improved. By October 2016, thirteen 24-hr sessions that targeted all
sources brighter than 100~mJy have been observed and analyzed. Among 3320 observed
target sources, 555 objects have been detected. We also conducted a number of
follow-up VLBI experiments in the refining mode and improved the positions of
249 ecliptic plane sources.

\end{abstract}


\section{Introduction} \label{sec:intro}

  One of the most important emerging Very Long Baseline Interferometry
(VLBI) applications is the phase referencing observations of spacecraft.
Such observations are used, for example, to provide wind measurements
of the atmosphere of Titan \citep{r:tit05}, to improve the Saturn
ephemerides by the astrometry of Cassini with the Very Long Baseline Array (VLBA)
\citep{r:cas15}, and to measure the lateral position of
the European Space Agency’s Mars Express (MEX) spacecraft
during its closest ever flyby of the Martian
moon Phobos \citep{2016A&A...593A..34D}. At the moment, there is the
Chang$'$e 3 lander on the Moon \citep{r:liu14,r:lic,r:zwm15} and in the near future
another lander, Chang$'$e 5, will be placed on the Moon. Phase
referencing observations of the radio beacon onboard these landers will
be used to study the Moon's libration and its response to solid tides induced
by the Earth and the Sun. Another important application is to observe Mars
landers, such as the InSight mission (2018) and the first Chinese Mars
mission (2020), for measuring the parameters of the Martian rotation.

  Nodding observations of a spacecraft and nearby calibrator radio sources
allow us to measure its position offset with a precision of up to several
tenths of a nanoradian (nrad)\footnote{1 nrad $\approx$ 0.2 milliarcsec (mas)}. One nrad
corresponds to $\sim$0.4~m on the Moon's surface and $\sim$100~m at the Martian orbit.
However, position offsets are useless for the scientific applications of
spacecraft observations. The goal of these observations is to determine
the absolute positions of a spacecraft. The error of the absolute position is
the quadrature sum of the error of the calibrator's absolute position and the error
of the position offset. The latter error largely depends on the angular distance
between the target and the calibrator. Hence, phase referencing observations
of spacecraft require a dense grid of calibrators along the ecliptic plane with
their absolute positions known with the highest possible accuracy.

We consider that a source is suitable as a calibrator
if its correlated flux density on the longest baseline of the array is
above the 30~mJy level. A 30~mJy source is detected reliably at two
antennas with the System Equivalent Flux Density (SEFD) 600~Jy for
50~s at a 2~Gbps data rate. Sometimes, when the received signals from a spacecraft
are very strong, we can use the spacecraft as a phase calibrator source, which
allows us to use much lower recording data rate.
In the past 15 years, over 19,000 sources were observed in a number of
dedicated astrometric VLBI observing campaigns \citep{r:vcs1,r:vcs2,r:vcs3,r:vcs4,
r:vcs5,r:vcs6,r:bessel,r:vips,r:vgaps,r:lcs1,r:egaps,r:obrs1,r:obrs2,
r:aofus2,r:aofus3}. A cumulative all-sky catalog of 12,651 sources called
rfc\_2016d (Petrov and Kovalev, in preparation) \footnote{Radio Fundamental Catalogue \url{http://astrogeo.org/rfc}} was
derived from the analysis of these observations. In this catalog,
the calibrator sources can be divided into three classes:
the 1st class, with position errors  of less than 1.5~nrad, suitable for
determining the absolute position of a target with the use of differential
astrometry, the 2nd class, with position errors of less than 10~nrad, suitable for
differential astrometry, and the 3rd class, with position errors of less than 100~nrad, suitable
for imaging. We select the threshold of 1.5~nrad for the 1st class calibrators,
because, above this level of position error, random measurement noise dominates.
Below that level, the main contribution comes from
systematic errors caused by mismodeled atmospheric path delay, source structure,
and the core-shift. Improvement of source position accuracy better than 1.5~nrad
requires significantly more observing resources.

The number of the 1st class calibrators in the ecliptic plane
is still not sufficient for the needs of space navigation or scientific applications.
There are two reasons: 1)~prior surveys were not deep enough and missed many suitable
weaker sources. 2)~the majority of these calibrators were derived from
very few observations. Many of them are made with a single band and,
therefore, their position estimates were affected by systematic errors
caused by the ionosphere.

  In 2015, we launched the VLBI Ecliptic Plane Survey (VEPS) program with
an ambitious goal to find {\bf all} suitable calibrators. Based on our
previous experience with running large VLBI surveys, we anticipated
a detection rate in the range of 15--30\%. Keeping this consideration in
mind, we run the program in two
modes: the search mode and the refining mode. In the search mode,
we observed all the targets with the minimum array
configuration. The objective of this step is just to detect all the sources
with correlated flux densities greater than 30~mJy, determine their coarse
positions with an accuracy of 10--50~nrad, and evaluate their average correlated flux
density. In the refining mode, we observe
in a large network all the sources detected in the search mode, as well as
previously known sources, with an integration time sufficient for deriving
their positions with an accuracy better than 1.5~nrad and generating their images.
The two-step approach optimizes resource usage by significantly reducing
time spent for observing the sources with emission from a compact region
that is too weak to be detected.

  In the search mode, {\it all} sources from the single-dish
Parkes-MIT-NRAO (PMN) \citep{1993AJ....105.1666G} and Green Bank 6~cm (GB6)
 \citep{1996ApJS..103..427G} catalogs within $7.5^\circ$
of the ecliptic plane brighter than 50~mJy
at 4.85~GHz are to be observed. Both input catalogs are considered complete
to that level of flux density. Observations in this mode are performed
at the X-band (8.2--9.0~GHz) only and use a 3--4 station network. In the refining
mode, observations are performed at the S/X (2.3/8.4~GHz) or C/X (4.3/7.6~GHz) dual
bands simultaneously with the use of a large network such as the VLBA, or the
International VLBI Service for Geodesy and Astrometry (IVS) network in a high
sensitivity mode. Here, we present the first results of the program derived
from observations of all sources brighter than 100~mJy in the search mode and
a number of sources observed in the refining mode.

\section{Source selection} \label{sec:source}

  We have selected all objects within $7.5^\circ$ of the ecliptic plane,
with single dish flux densities brighter than 50~mJy at 5~GHz from
the PMN and GB6 catalogs, except those that a)~have been detected with VLBI
before and b)~were observed with VLBI in a high sensitivity mode (detection
limit better than 20~mJy), but too weak to be detected. Contrary to many prior
surveys, we did not preselect targets based on source spectral index, since
almost all the sources with flat spectra have already been observed.
The PMN catalog has two small zones, located at $200^\circ$ ecliptic longitude,
which misses sources due to solar contamination when
the sidelobes of the antenna encountered the Sun, so those data
have been expunged from the survey \citep{1995ApJS...97..347G}. We included in these zones
789 sources from the NRAO VLA Sky Survey (NVSS) catalog \citep{r:nvss}
that are brighter than 50~mJy at 1.4 GHz.

  In total, there are 7807 target sources in our list. Of those,
approximately 1/3 have flux densities above 100~mJy and 2/3 have
flux densities in the range 50 to 100~mJy. These flux densities were
measured with a single dish telescope or connected interferometers
with beam size $40''$--$200''$. Emissions
from scales of 5--50~nrad detectable with VLBI represent only a fraction of
the total emission at arcminute scales. Therefore, a number of sources
are expected to have correlated flux densities below the detection limit.

  We organized observations in such a way that the stronger
sources were observed first and weaker sources were observed later.

\section{Observations} \label{sec:obs}

\subsection{Observations in the search mode} \label{sec:obs_sea}

  We began observations in the search mode in February 2015.
The participating stations included the three core stations of the Chinese
VLBI Network (CVN): {\sc seshan25}, {\sc kunming}, and {\sc urumqi}.
However, sometimes these stations were not available at the same time,
or occasionally one of them had a risk of failure. In that case, one
or two international stations joined. Depending on the participating
stations, the longest baseline length in each session can be varied
from 3200~km to 9800~km.

Figure~\ref{map} shows the geographical distribution of all
participating stations. {\sc kashima34}, {{\sc sejong} and {\sc hobart26}
have contributed to past VEPS observations. They are relatively large
antennas, and have good common visibility of the ecliptic zone. Before
joining in the VEPS survey, we performed fringe tests to {\sc sejong},
{\sc hobart26}, and {\sc kashim34} in December 2014, July 2015, and January
2016, respectively.

\begin{figure}[htb!]
  \includegraphics[width=.45\textwidth]{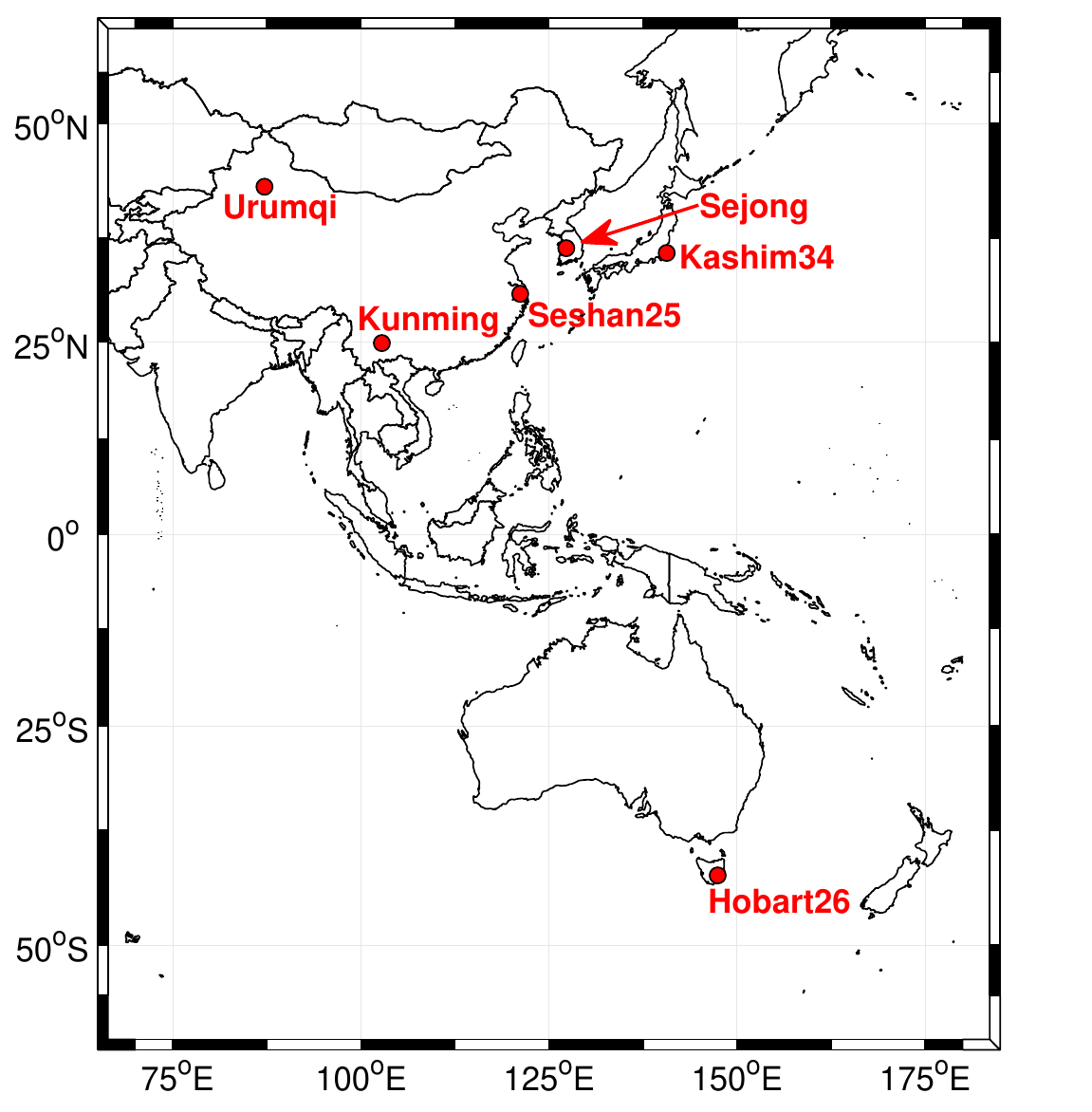}
  \caption{Distribution of participating stations.}
  \label{map}
\end{figure}

Our observations were performed at a 2048~Mbps data rate with
16 Intermediate Frequency (IF) channels and 2-bit sampling.
The first eight IFs of 32~MHz bandwidth were distributed in the
range of [8.188, 8.444]~GHz, and the remaining eight IFs
of 32~MHz bandwidth were spread in the range of [8.700, 8.956]~GHz.
The data volume was around 16~TB for each station in one 24-hr session.

We used the Chinese VLBI Data Acquisition System (CDAS)
for data acquisition at the Chinese stations,
ADS3000+ at {\sc kashim34}, and K5/VSSP32 for {\sc sejong}.
The maximum data rate of the K5/VSSP32 is limited to 1024 Mbps, so the data
were sampled with 1 bit. For {\sc hobart26},
the Digital Base-band Converter (DBBC) board was used.
At the time, the firmware supported only 16~MHz-wide IFs,
so we observed at 16~MHz-wide IFs instead and correlated them against
the low parts of 32~MHz-wide IFs recorded at other stations.

  Each target source was observed in two scans of 90 s length, with a gap
between consecutive observations of the same source of at least 3 hr.
The sequence of observations was optimized to minimize slewing time with
the use of the astrometry VLBI scheduling software package {\sf sur\_sked}.
Every 1 hr, four calibrators from the pool of 190 bright sources were inserted
in the schedule. Two of them were observed at elevations below $30^\circ$ and
two at elevations of $50^\circ$ or more above the horizon. They were also used
for bandpass calibration, for antenna gain calibration, as ties for
astrometric global solutions, and for improving the estimation of the
atmosphere path delay in the zenithal direction.

\begin{deluxetable}{ccclc}[h!]
\tablecaption{Summary of the VEPS observations in the search mode.
\label{tab:sessions}}
\tablecolumns{12}
\tablewidth{0pt}
\tablehead{
\colhead{Date} &
\colhead{Dur.} & \colhead{Code} & \colhead{Stations\tablenotemark{a}} & \colhead{\# Targets} \\
\colhead{(YYYY-mm-dd)} & \colhead{(hrs)} &
}
\startdata
2015-02-13   & 24 & VEPS01 & ShKmUr    & 293   \\
2015-02-14   & 24 & VEPS02 & ShKmUr    & 338   \\
2015-04-23   & 24 & VEPS03 & UrKv      & 300   \\
2015-04-24   & 24 & VEPS04 & ShKmUrKv  & 400   \\
2015-08-10   & 25 & VEPS05 & ShKmKvHo  & 252   \\
2015-08-19   & 25 & VEPS06 & ShKmKvHo  & 277   \\
2016-03-02   & 24 & VEPS07 & ShKmUrKb  & 333   \\
2016-03-11   & 24 & VEPS08 & ShKmUrKb  & 477   \\
2016-05-13   & 24 & VEPS09 & ShUrHo    & 291   \\
2016-05-14   & 22 & VEPS10 & ShUrKv    & 322   \\
2016-07-06   & 24 & VEPS11 & ShUrKb    & 307   \\
2016-09-02   & 23 & VEPS12 & ShUr      & 424   \\
2016-09-03   & 23 & VEPS13 & ShKmUr    & 344   \\
\enddata
\tablenotetext{a}{
Sh: {\sc Seshan25}; Km: {\sc Kunming}; Ur: {\sc Urumqi}; Kv: {\sc Sejong}; Kb: {\sc Kashim34}; Ho: {\sc Hobart26}.
}
\end{deluxetable}

  By September 2016, 13 sessions had been observed. The summary of these
observations is presented in Table~\ref{tab:sessions}. In general, the
observations were successful, despite a number of failures. {\sc urumqi} had
a receiver problem in the first 12 hr in VEPS01. {\sc seshan25} and
{\sc kunming} made use of wrong B1950 source positions for antenna control in VEPS03.
{\sc hobart26} had a timing problem after 9 hr in VEPS09. {\sc kunming}
data in VEPS012 were lost due to a hard disk failure. If a target
source was not observed due to station failure, we reobserved it in the next
VEPS sessions.

\subsection{VLBA observations in the refining mode} \label{sec:obs_vlba}

  We ran two absolute astrometry dual-band VLBA programs that targeted ecliptic
plane compact radio sources: the dedicated survey of weak ecliptic plane
calibrators with the VLBA, BS250 program in March--May 2016,
and the VLBA Calibrator Survey~9 (VCS--9)
in August 2015--September 2016.

  The BS250 program had 111 target sources within $7.5^\circ$ of the ecliptic
plane observed in four 8-hr segments. The targets were the weakest calibrators,
with correlated flux densities at baseline projection lengths greater than 5000 km in
the range [30, 50]~mJy. Each target was observed in three scans of 180~s length.
The target sources were scheduled in such a way that the minimum gap between their
consecutive observations was at least 2 hr. Every 1.5~hr a block of four
calibrators from the  pool of bright sources was inserted in the schedule in
such a way that two of them were observed at elevations below $30^\circ$ and
two at elevations $50^\circ$ or more above the horizon, similar to the VLBI
observations in the search mode. We used the same frequency setup for the BS250 as
in the VCS-II program \citep{2016AJ....151..154G}: four IFs of 32~MHz
bandwidth spread in the range of [2.22, 2.38]~GHz and 12 IFs spread in the range of
[8.43, 8.91]~GHz recorded simultaneously at a 2~Gbps data rate. Of the 111 target
sources, 37 were detected in the search mode of the VEPS program and the reminder
were detected in other surveys.

  The VCS--9 program has a goal of the densification of the grid of VLBI calibrators,
and observed over 11,000 sources spread approximately uniformly over the sky at
declinations above -$45^\circ$ in one scan of 60~s each. There is some overlap
between the source lists of VCS--9 and VEPS. The position accuracy of
the sources observed in VCS--9 is worse than those observed in the BS250 campaign, since
its integration time was nine times less, but is still significantly higher than in
the VEPS search mode. Therefore, we consider the VCS--9 program to be observations
in the refining mode for the purposes of this study. By February 2017, about 1/2
of the VCS--9 observations had been processed, so we report here only the VCS--9 results
available so far of the sources detected in the VEPS search mode. VCS--9
observed with the wide-band C-band receiver and recorded eight~IFs, 32~MHz-wide,
spread within [4.13, 4.61]~GHz, and eight IFs spread within [7.39, 7.87]~GHz
simultaneously. Scheduling VCS--9 observations was done in a similar way to those of BS250,
except only one scan per source was observed and the interval of time between
calibrators was reduced to 1 hr. VCS--9 was observed in segments of
3.5--10 hr long.

\subsection{IVS observations in the refining mode} \label{sec:obs_ivs}

 The IVS runs a number of VLBI observing programs primarily for geodesy with
occasional use for astrometry \citep[e.g.][]{r:bail}. As a subgroup of the IVS,
the Asia-Oceania VLBI network (AOV), that includes facilities from five
countries in the Asia-Oceania region: Australia, China, Japan, New
Zealand and South Korea \citep{r:lovell15}, ran a number of experiments
beginning in 2015. We made an attempt to improve the coordinates of
some VEPS sources detected in the search mode and provide additional
measurements of telescope
position with the same experiments in two such 24-hr sessions,
AOV010 in July and AUA012 in August 2016.

  Both sessions included sensitive AOV antennas: {\sc parkes} and
{\sc hobart26} in Australia; {\sc seshan25}, {\sc kunming} and {\sc urumqi}
in China. In addition, {\sc tianma65} (Tianma 65m Radio Telescope, or TMRT)
in China and {\sc tsukuba} in Japan
participated in AOV010. We conducted observations at a 1~Gbps data rate with 16 IFs
of 16~MHz bandwidth and 2-bit sampling, which is the highest data rate that all of the
participating stations were able to provide in 2016.

  Since two geodetic stations, {\sc kashim11} and {\sc koganei} could observe
only within a narrow X-band (8.1--8.6~GHz), we adjusted the frequency sequence
by balancing the uncertainty of the group delay and the amplitude of the highest
side lobe in the delay resolution function. The following frequency sequence was
used in the session AOV010: [8.19799, 8.21399, 8.23399, 8.25399, 8.33399,
8.41399, 8.51399, 8.53399, 8.55799, 8.57399]~GHz. The amplitude of the
highest side lobe is 0.52.

  We used geodetic software {\sc sked} \citep{r:sked} for scheduling these
experiments which is widely used in the preparation of many other IVS sessions.
We set a group of weak sources as targets. Among them, 32 sources were previously
detected in the VEPS search mode observations, with declinations in the
range [-$31^\circ$, -$15^\circ$]
and correlated flux densities in the range [30, 80]~mJy. Besides
that, a group of strong compact sources were selected automatically in order
to provide a uniform sky coverage. We used the astrometric mode of
{\sc sked} for automatic scheduling, which is described in more details in
\citet{r:bail}. After taking a few iterations by adjusting some control
parameters of the {\sc sked}, we were able to schedule 60\% of the total
number of observations for target sources,
while keeping enough scans for small antennas to achieve the geodesy goals.

\section{Data processing}

\subsection{Data correlation}

   We correlated the VEPS search mode observations and the IVS sessions
with the DiFX software correlator \citep{r:difx2}, which was installed on a powerful
hardware platform in 2014 \citep{r:coshao} at the Sheshan Campus of the Shanghai
Astronomical Observatory. The data from the Chinese domestic stations were
recorded on disk packs and then shipped to Shanghai, while the data from
international stations were transferred to Shanghai via a high speed network.
The data volume for each station is approximately 16~TB, eight times bigger than
that recorded in regular IVS geodetic sessions, so the data correlation of
one VEPS session usually took more than 24 hr.

Correlation of mixed observing modes with different bandwidths or sampling bits
can be challenging. Fortunately the DiFX can fully support the correlation of 1-bit
sampled data from {\sc sejong} against 2-bit sampled data from the other stations.
For the correlation of 16~MHz bandwidth data from {\sc hobart26} against
the 32~MHz bandwidth data from the other stations, the zoom mode was selected to pick up
the overlapped frequency band. Moreover, it was optional to make
correlations only on the 16~MHz bandwidth on the baselines to {\sc hobart26},
while the other stations with 32~MHz bandwidth went through an
independent correlation pass, the same as the usual correlation
procedures. We selected an accumulation period of 0.125~s,
and 512 spectral channels per IF. This setup provides
a wide field of view that allowed us to detect a source within
several arcminutes of their a~priori position. In fact, we observed
fields around the pointing direction rather than individual sources.

  Correlation of the VLBA experiments was performed at the Socorro array control
center using the same DiFX software correlator. The DiFX correlator output
was converted into FITS-IDI format using the difx2fits program for further
analysis.

\subsection{Post-correlation analysis}

   The visibility data produced by the correlator were then processed using
the VLBI data analysis software \PIMA\hspace{-0.2em}\footnote{See \href{http://astrogeo.org/pima}
{http://astrogeo.org/pima}}. A detailed description of the analysis strategy
and a comparison between the methods adopted in the past and those used for
processing our data can be found in \citet{r:vgaps}. We ran the pipeline
that includes a coarse fringe search, manual adjustment of phase
calibration, generation of the auto-correlation and cross-correlation bandpass
masks, computation of the complex bandpass calibration, a fine fringe search,
amplitude calibration, amplitude normalization and computation of the
total group delay. The group delays, fringe amplitudes and related information
were written in databases for further analysis.

  The databases were loaded into the VTD/Post-Solve software\footnote{See \href{http://astrogeo.org/vtd}{http://astrogeo.org/vtd}}
for preliminary astrometric data analysis. At the beginning of the astrometric
analysis, we discarded observations with a signal to noise ratio (SNR) of
less than six, which corresponds to the probability of false detection
around 0.001. The SNR is defined here as the ratio of the peak amplitude
to the mean amplitude of the noise. In the beginning, we used a coarse parameter
estimation model: we estimated only the positions of the sources not detected
before with VLBI and a clock function that is represented by a linear
spline with knots placed at equal intervals of 1~hr. After the elimination of
outliers, we refined the model by adding an estimation of source positions and
residual atmosphere path delay in the zenithal direction. Then we reduced the SNR
limit gradually to 5.2. We discarded all new sources with less than two detections.

  Then we updated the source positions, reran fringe fitting and repeated
the procedure of astrometric data analysis. For those
sources that were marked as outliers after the second step of iterations,
we computed the expected group delay based on results of parameter estimation.
Then we reran fringe fitting for these observations with
a narrow fringe search window and repeated the astrometric analysis.

Databases cleaned for outliers were saved for the final astrometric analysis.

\subsection{Astrometric analysis}

  We ran an astrometric analysis in the global mode, which is the usual
approach for processing absolute astrometry VLBI surveys \citep{r:vcs6}.
That means we used all VLBI experiments in geodesy and absolute astrometry
acquired so far, and the new data. We ran two solutions. In solution A we used
all geodesy data acquired until December 1, 2016, absolute astrometry VLBI
data acquired until January 1, 2015 (before the start of the VEPS program)
and the VEPS--1 data in the search mode. In solution B we used all dual-band
VLBI data acquired until February, 2017, including the BS250 and VCS--9 campaigns,
but excluding single-band group delays acquired in the search mode of the VEPS
program.

  We estimated source coordinates, station positions and velocities as global
parameters; pole coordinates, UT1 angle, their rates, and daily nutation offsets
for every observing session; clock function and residual zenith path delay
in the atmosphere modeled with a B-spline of the first degree with a step of
60 and 20~min respectively. No-net-rotation constraints were applied to the
estimates of the source coordinates in such a way that the net rotation of 212
sources marked as ``defining'' in the ICRF catalog \citep{r:icrf98} was zero
with respect to their catalog positions. The data analysis
procedure is very close to that used for deriving other VLBI catalogs,
e.g. in \citet{r:obrs2}.

We added in quadrature a floor of 0.5~nrad to the position uncertainties computed
with the law of error propagation in order to accommodate the contribution of
unaccounted systematic errors. The value of the floor was found empirically from the set of trial solutions and comparisons of the source coordinate estimates from independent subsets of observations.
The uncertainties inflated with the given floor are in a closer agreement with the differences in source coordinate estimates from VLBA observations.

  Since VEPS search mode experiments were observed in the single band,
we computed the ionosphere contribution from the total electron content (TEC)
maps provided by the CODE analysis center for processing Global Navigation
Satellite System data \citep{r:scha98} using the technique described in detail
in \citet{r:vgaps}. Our previous extensive analysis of the residual contribution
of the ionosphere at 8~GHz to the source coordinate estimates after applying
the a~priori path delays computed from TEC maps to the data reduction model
does not exceed 7.5~nrad in quadrature in the worst case \citep{r:lcs1}.
Since uncertainties of source positions derived from VEPS search mode observations
are greater than that, we just ignored the contribution due to errors in the
TEC model.

\subsection{Imaging analysis of VLBA observations }

  Using the results of fringe fitting, we applied the \PIMA\ task {\sf splt}
to perform coherent averaging over time and frequency after phase rotation
according to group delays and phase delay rates, apply calibration for
system temperature, gain curves, bandpass renormalization, combine all
visibilities of a given source, and write averaged visibilities and their
weights into output binary files in the FITS format. The data were then suitable for
imaging with the NRAO Astronomical Image Processing
System ({\sc AIPS}) and the Caltech {\sc difmap} package \citep{Shepherd94}.

  We ran this procedure through all segments of the BS250 campaign. Some sources
were observed in more than one segment and, since the observations took place
within two months, we ignored possible source variability and merged the
calibrated visibilities.

  The visibilities coherently averaged over all spectral channels within
an individual IF and over time with an integration time of 8 s. The resulting
data were exported to
{\sc difmap} for imaging and calibrating the residual phase errors. We performed
a traditional hybrid mapping procedure consisting of several iterations of
{\sc clean}ing \citep{Hogbom74} and phase and amplitude self-calibration.
We first self-calibrated the phases against a model of point-like sources using the
{\sc startmod} task. Then we performed phase self-calibration and mapping
under uniform weighting, followed by natural weighting. Every source was
first imaged with an automatic pipeline. The results were examined and those
sources for which the automatic pipeline did not provide satisfactory results
were reimaged manually.
The main reason for manual reimaging was the necessity to remove
outliers due to inaccurate on-off source flagging, and due to spikes in system
temperature readings caused by radio frequency interference (RFI).
Thus the data were edited and those parameters including
clean boxes were set by hand in order to avoid the big sidelobes and get
more correct morphology.
Figure~\ref{fig:f2} shows
some sample images. The final images in FITS format, as well as self-calibrated
visibilities, are accessible from the project web site
\url{http://astrogeo.org/veps}.

\begin{figure*}
\plottwo{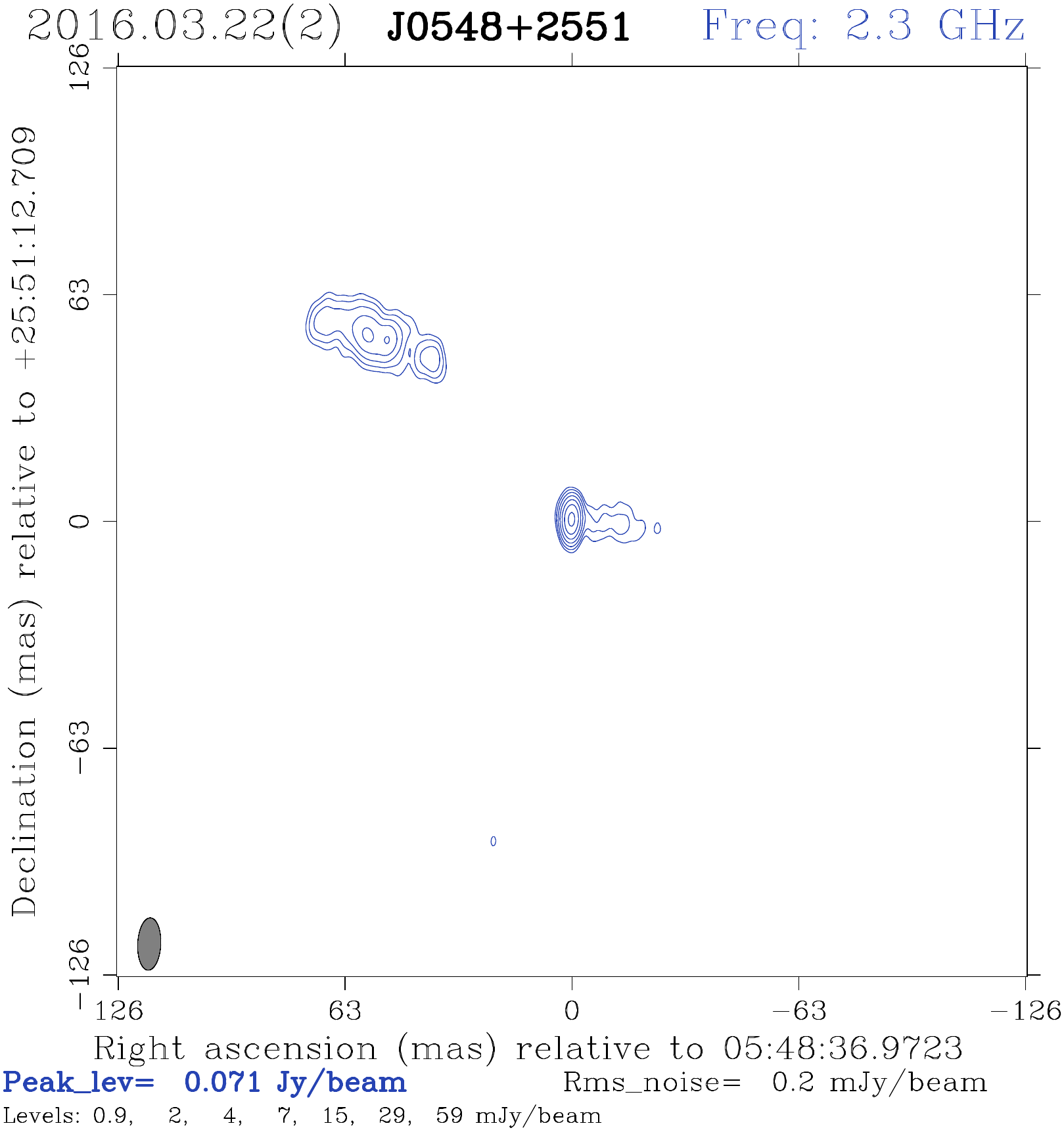}{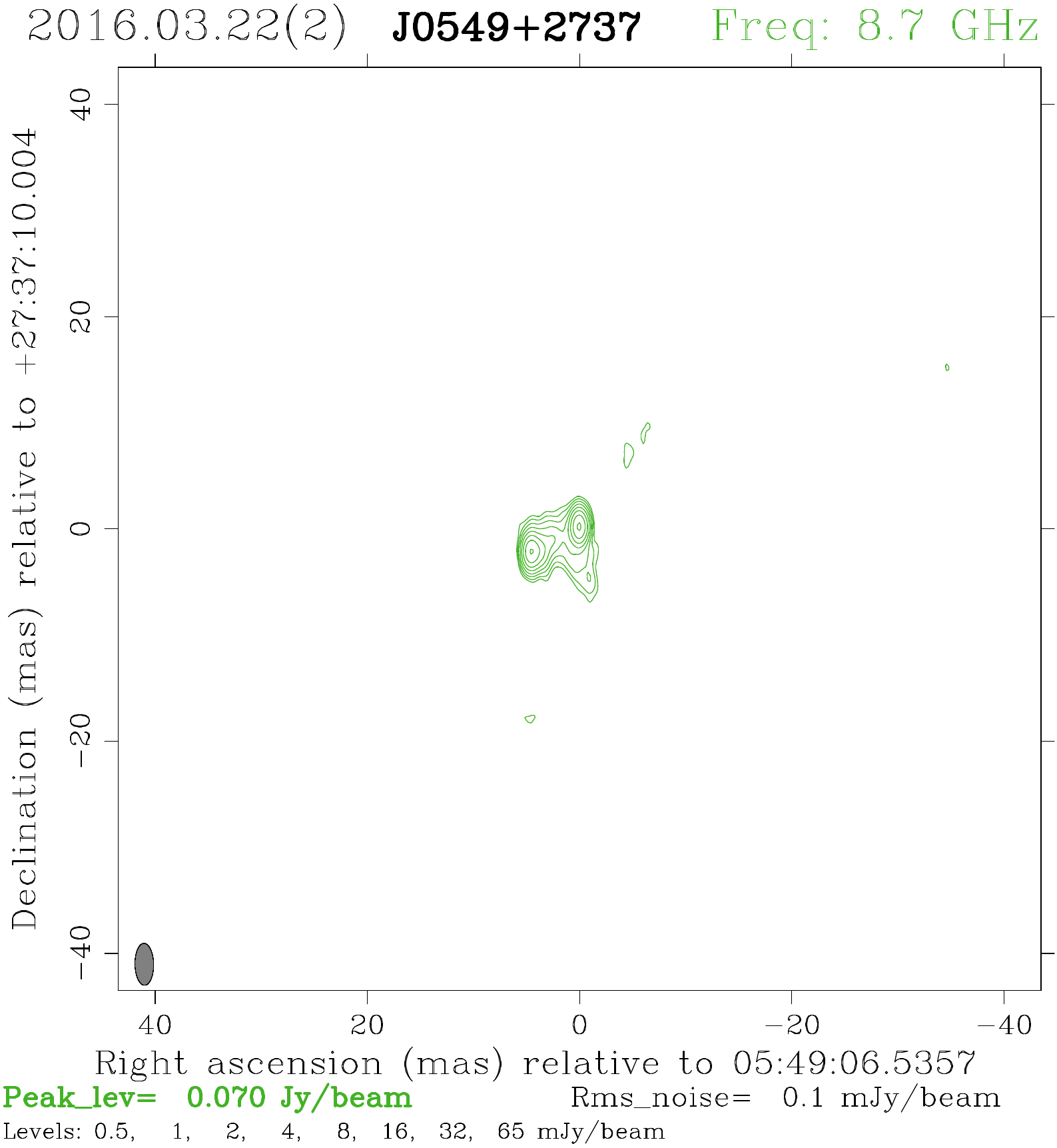}
\caption{Left: the image of J0548+2551 at S-band. There is a counter-jet and
some extended emission far away from the central core.
Right: the image of J0549+2737 at X-band. This CSO (Compact Symmetric Object)
has a significant structure effect.
The lowest counter was set at three times the root-mean-square (rms)
noise of the residual image. \label{fig:f2}}
\end{figure*}

  Images are necessary for planning phase referencing observations in order to
predict the necessary integration time. We are going to reanalyze the data with
the source structure contribution applied in the future in order to improve
position accuracy and to tie the reported position to a particular source image
feature.

\subsection{Amplitude analysis in VEPS--1 experiments}

   Since the VEPS search mode sessions had too few observations per source,
typically 2--6, imaging was not feasible. Instead, we used the non-imaging
procedure developed in \citet{r:lcs1} for computing the mean correlated flux
density. Firstly, we calibrated the raw visibilities data for the a~priori
antenna gain $G(e)$ and system temperature $T_{sys}$:
$ F_{corr} = v \cdot T_{sys}(t,e)/G(e)$. Secondly, we adjusted antenna gains
using publicly available brightness distributions of the calibrator sources made
with observations under other programs that can be found in the Astrogeo
VLBI FITS Image Database\footnote{\href{http://astrogeo.org/vlbi\_images}{http://astrogeo.org/vlbi\_images}}.
Using images in the form of CLEAN components, we computed the predicted flux densities
of the calibrator sources for every observation. Using logarithms of the flux densities
of the calibrator sources derived from the calibrated visibilities and from known
images, we computed the multiplicative gain corrections using least squares.
These gain corrections were applied to the target sources and the correlated
flux density estimates were corrected for errors in a~priori gain
calibration. This procedure is known to have errors of around 15\%
\citep{r:lcs1,r:qcal1}.

\section{Results}

\begin{figure}
    \includegraphics[width=0.45\textwidth]{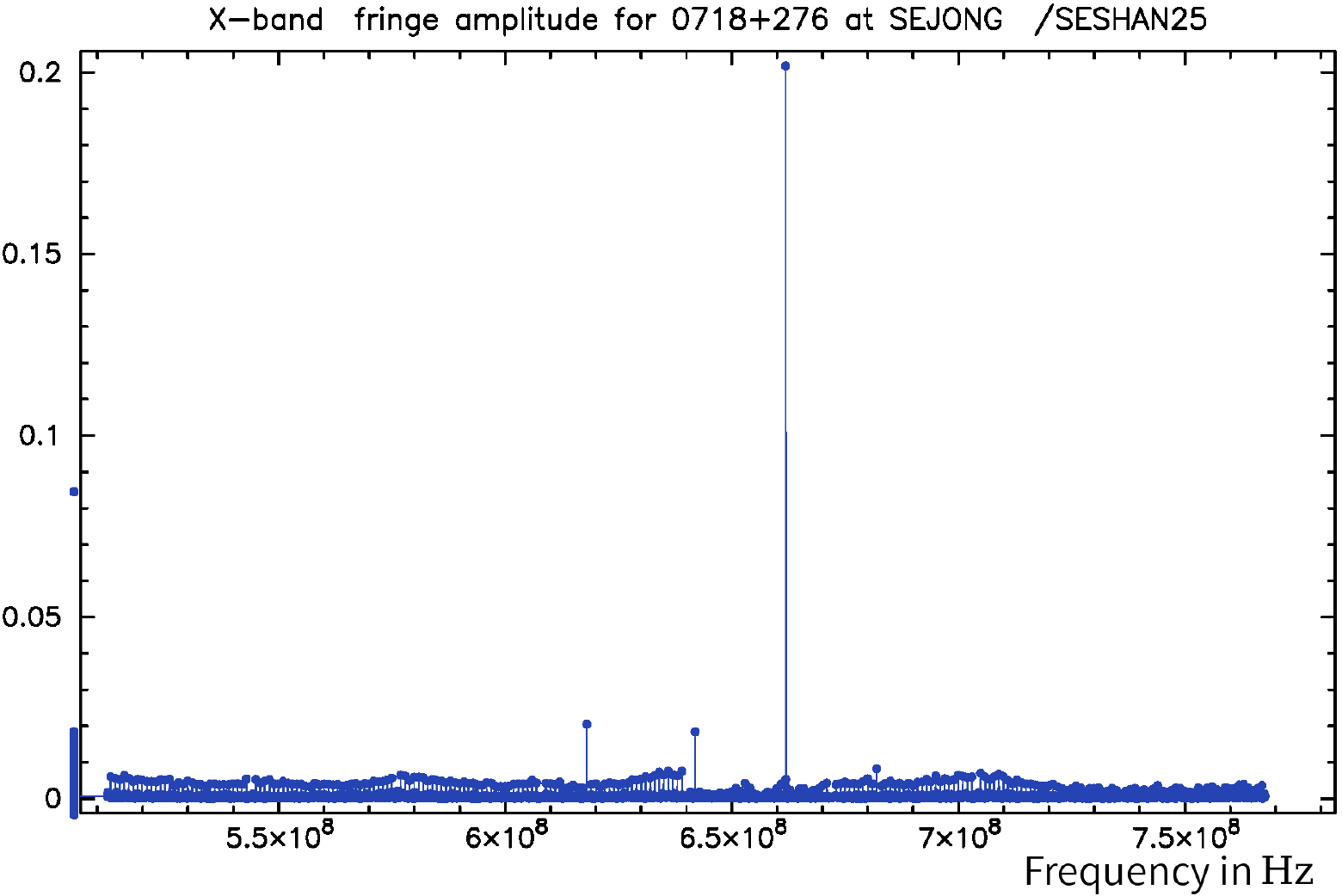}
    \caption{Dependence of uncalibrated fringe amplitude on frequency
             for an observation affected by radio interference.
             A portion of the spectrum relative to the reference frequency
             8.188~GHz is shown.
            }
    \label{f:rfi_frq}
\end{figure}

\begin{figure}
    \includegraphics[width=0.45\textwidth]{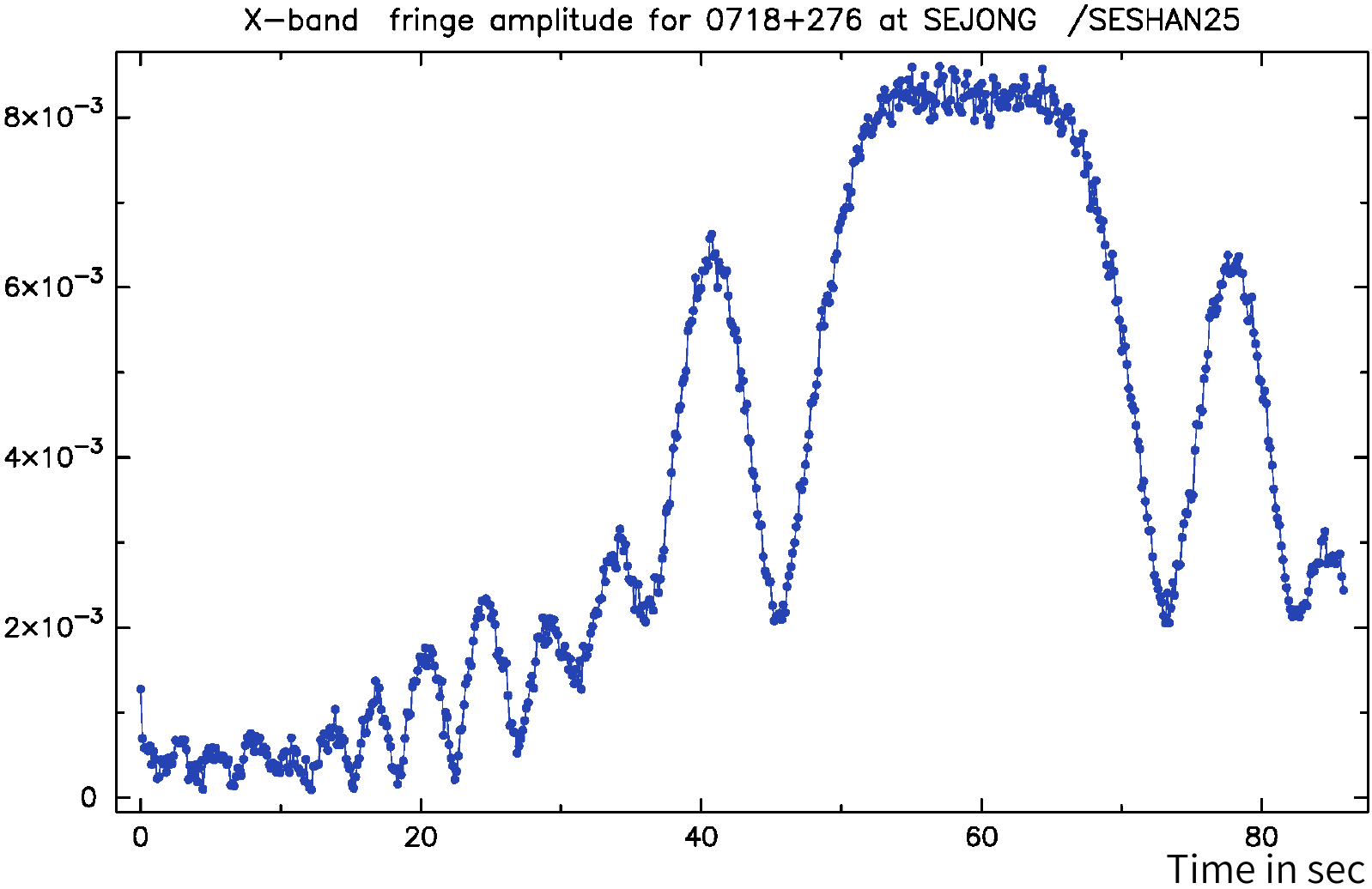}
    \caption{Dependence of uncalibrated fringe amplitude on time
             for an observation affected by radio interference.
            }
    \label{f:rfi_tim}
\end{figure}

\begin{table*}[!htb]
    \caption{The first 4 rows of the table of 555 target sources detected in the
             search mode of the survey. Column 1 contains a flag of the follow-up
             observations: V for VCS--9, B for BS250, and I for IVS. Column 6--9 contain
             the uncertainty in right ascension without $\cos\delta$ factor,
             the uncertainty in declination, the correlation between right ascension and declination estimates and the number of observations used in the solution.
             Column 10 contains the estimate of the median flux density at 8~GHz.
             The full table is available in the electronic attachment.}
    \par\medskip\par
    \centering
    \begin{tabular}{rrrrrrrrrr}
       \hline
         & \nntab{c}{IAU source name} & \nntab{c}{J2000.0 source coordinates} &
           \nnntab{c}{Position errors} & & \\
       F & B-name     & J-name       & Right ascension & Declination     & $\sigma_\alpha$ & $\sigma_\delta$ & Corr & \# obs & $F_{\rm med}$ \\
         &            &              & h \hspace{0.1em} m\hspace{0.4em} s \hspace{2.4em} \phantom{a} &
                                     ${}^\circ \hspace{1em} ' \hspace{1em} '' \hspace{2.5em}$ &
                                     mas & mas & & & Jy \\
       \ntab{c}{(1)} & \ntab{c}{(2)} & \ntab{c}{(3)} & \ntab{c}{(4)} & \ntab{c}{(5)}  &
       \ntab{c}{(6)} & \ntab{c}{(7)} & \ntab{c}{(8)} & \ntab{r}{(9)} & \ntab{c}{(10)} \\
       \hline
         & 2358$-$072 & J0001$-$0656 & 00 01 25.586899 & $-$06 56 24.93216 &  1.56 &  3.24 &  0.052 &   6 & 0.051 \\
       V & 2359$-$038 & J0002$-$0331 & 00 02 30.622538 & $-$03 31 40.45732 &  4.14 & 10.57 &  0.239 &   5 & 0.022 \\
         & 0000$-$044 & J0002$-$0411 & 00 02 41.255161 & $-$04 11 55.30522 &  2.27 &  3.94 &  0.596 &   6 & 0.023 \\
         & 0000$-$006 & J0002$-$0024 & 00 02 57.175395 & $-$00 24 47.27274 &  2.33 &  5.98 &  0.841 &   4 & 0.035 \\
       \ldots &&&&&&&&& \\
       \hline
    \end{tabular}
    \label{t:cat1}
\end{table*}

  We found 520 sources in the VEPS--1 program with three or more good
observations and 51 sources with two observations out of 3320 target sources
observed. The minimum number of observations for the determination of source
coordinates is two. However, if only two observations are used for deriving
source positions and one of them is bad, i.e. either affected by the RFI
or affected by a failure in the fringe fitting process, such an
error will not be noticed but can shift the estimate of source position by
a large amount, up to several arcminutes. Three good observations used in the least
square solution provide the minimum redundancy and greatly reduce the probability
that the source position is affected, to a non-negligible level, by an unnoticed 
failure in group delay determination. However, there were two experiments in
the VEPS search mode that had usable data from only a single baseline because
of station failures. Therefore, we examined all observations of the sources
that had only two detections. We discarded observations that had a
SNR $<$ 7, which is about 20\% above the detection limit, and then manually
screened fringe plots for abnormalities among the remaining observations.
Usually, sources of RFIs are narrow-band and have a terrestrial origin.
Therefore, their spectrum has sharp peaks and the fringe amplitude has strong
dependence on time, since phase rotation that compensates for the Earth's rotation
was added during correlation. Figures~\ref{f:rfi_frq}--\ref{f:rfi_tim}
illustrate the dependence of fringe amplitude on frequency and time of
observation affected by the RFI. This dependence is supposed to be flat for
normal observations. Since in that case the peak of the fringe amplitude was
exactly at 8.850~GHz, we conclude that the peak was caused by local
interference generated by the VLBI hardware.
We analyzed the pattern of fringe phase and fringe amplitudes and
removed from the dataset the observations with similar pattern.
If a source was observed in only two scans, the dataset has no
redundancy and when we estimate source coordinates, unlike
the case when there are three or more observations,
the residuals will be very close to zero even for observations
with group delay severely corrupted by the RFI.
We kept 35 sources detected with only two observations
with SNR $> 7$ and without abnormalities in their fringe plots,
though there is a risk of large offset for the position estimates
of a few sources.

   The SEFD at stations {\sc urumqi}, {\sc seshan25}, {\sc kunming}, and
{\sc kashima} in elevations 20--$90^\circ$ was in the range of 300--800~Jy.
The detection limit at the baselines with these sensitive antennas was
in the range of 13--18~mJy. The SEFD at {\sc hobart26} was
in the range of 1300--1800~Jy and at {\sc sejong} was in the range
of 3000--5000~Jy. The detection limit at the baselines with {\sc hobart26}
or {\sc sejong} was in a range 30--60~mJy. All the sources,
except 51 mentioned above, have at least three
observations at sensitive baselines. Therefore, we conclude that we have detected
all target sources with a correlated flux density greater than 20~mJy.
The detection rate was 19\% for the sources with galactic latitude
$|b|>10^\circ$ and almost two times less, 8.5\%, for the sources with galactic
latitude $|b|<10^\circ$.

\begin{table*}[ht]
    \caption{The first 4 rows of the table of 249 sources observed in the
             refining mode of the survey. Column 1 contains a flag of the follow-up
             observations: V for VCS--9, B for BS250, and I for IVS. Column 6--9 contain
             the uncertainty in right ascension without $\cos\delta$ factor,
             the uncertainty in declination, the correlation between right ascension and declination estimates and the number of observations used in the solution.
             Column 10--15 contain the estimate of flux density at X~band
             (8.4 or 7.6~GHz), C~band (4.3~GHz), and S~band (2.3~GHz). Two estimates per
             band are provided: the total flux density integrated over the map and
             the unresolved flux density computed as the median flux density at
             baseline projected lengths over 5000~km. The full table is available in
             the electronic attachment.}
    \footnotesize
    \par\medskip\par
    \begin{tabular}{r@{\quad} l@{\enskip} l@{\enskip} r@{\enskip} r@{\quad} r@{\enskip}
                   r@{\enskip} r@{\enskip} r@{\quad} r@{\enskip} r@{\quad} r@{\enskip}
                   r@{\quad} r@{\enskip}r}
       \hline
         & \nntab{c}{\ns IAU source name} & \nntab{c}{\ns J2000.0 source coordinates} &
            \nnntab{c}{\ns Position errors} & & \nnnnntab{c}{\ns Flux density estimates} \\
       F & B-name     & J-name     & Right ascension & Declination     & $\sigma_\alpha$ & $\sigma_\delta$ & Corr & \#obs
                      & $F_{\rm x,tot}$ & $F_{\rm x,unr}$ & $F_{\rm c,tot}$ & $F_{\rm c,unr}$ & $F_{\rm s,tot}$ & $F_{\rm s,unr}$ \\
         &            &              & h \hspace{0.1em} m\hspace{0.4em} s \hspace{2.4em} \phantom{a} &
                                     ${}^\circ \hspace{1em} ' \hspace{1em} '' \hspace{2.5em}$ &
                                     mas & mas & & & Jy & Jy & Jy & Jy & Jy & Jy \\
       \ntab{c}{(1)} & \ntab{c}{(2)} & \ntab{c}{(3)} & \ntab{c}{(4)} & \ntab{c}{(5)} &
       \ntab{c}{(6)} & \ntab{c}{(7)} & \ntab{c}{(8)} & \ntab{c}{(9)} & \ntab{c}{(10)} &
       \ntab{c}{(11)} & \ntab{c}{(12)} & \ntab{c}{(13)} & \ntab{c}{(14)} & \ntab{c}{(15)} \\
       \hline
       B & 2358$-$080 & J0001$-$0746 & 00 01 18.024906 & $-$07 46 26.92254 &  0.28 &  0.58 &  0.069 &  89 & 0.180 & 0.073 &       &       & 0.214 & 0.111 \\
       V & 2359$-$038 & J0002$-$0331 & 00 02 30.622737 & $-$03 31 40.44165 &  1.84 &  5.01 & -0.029 &  19 & 0.022 & 0.023 & 0.026 &       &       &       \\
       V & 0002$-$018 & J0005$-$0132 & 00 05 07.071782 & $-$01 32 45.12985 &  0.65 &  1.60 & -0.157 &  35 & 0.042 & 0.026 & 0.056 & 0.025 &       &       \\
       B & 0007$+$016 & J0009$+$0157 & 00 09 58.657674 & $+$01 57 55.14930 &  0.24 &  0.47 &  0.050 & 161 & 0.092 & 0.039 &       &       & 0.153 & 0.042 \\
       \ldots &&&&&&&&&&&&&& \\
       \hline
    \end{tabular}
    \label{t:cat2}
\end{table*}

 Table~\ref{t:cat1} presents the VEPS--1 catalog of 555 target sources from the
astrometric solution~A. The first column contains a flag that shows whether a given
source was observed in the VCS--9, the BS250 or the IVS
campaigns, but whose positions were derived solely from VEPS search mode observations.
The semimajor axis of the error ellipse ranges from 3.2 to 648~nrad, with median value
of 20.7~nrad. For almost all the sources, the thermal errors dominate the systematic
errors induced by the residual ionosphere. As a measure of source brightness, we use
the median correlated flux density. The correlated flux density can vary by
more than one order of magnitude for a resolved source with core-jet
morphology, depending on the baseline vector projection. Therefore, the median
flux density provides the upper limit of the unresolved flux density and should
be used with care. The median flux density of the VEPS--1 catalog varies
in the range of 0.013 to 0.34 Jy with the median 0.051~Jy.

   Table~\ref{t:cat2} presents the positions of the 249 objects derived from
the VLBA and IVS observations of the sources within $7.5^\circ$ of the ecliptic
plane. Their coordinates were estimated in the solution~B.
It should be noted that 4 weak sources observed in VCS--9 have position errors
larger than 70 nrad due to limited observations. The semimajor axis of the error ellipse for
the remaining 245 sources ranges from 0.5 to 46~nrad, with the median value of 3.0~nrad,
which is a factor of seven less than the position errors derived from the analysis of VEPS
observations in the search mode.
The last six columns contain estimates of flux
densities at 8, 4, and 2.3~GHz, the total flux density computed by the integration
of all CLEAN components in the image, and the unresolved flux density, defined as
the median flux density at baseline projection lengths greater than 5000~km.
The correlated flux density at a given baseline projection vector will be
within the range of the unresolved and the total flux density. These flux density
estimates were derived from source images. The images used for the generation of
the estimates of flux densities are publicly available from the Astrogeo VLBI
FITS image database.

\section{Discussion}

   The distribution of the undetected and detected sources observed in the search mode
is presented in Figure~\ref{f:veps1_distr}. All the target sources
with single-dish flux densities at 5 GHz greater than 100~mJy have been observed,
except for the zone within galactic latitude
$|b|<5^\circ$ in the range of right ascensions of 17--19~hr.
There are two reasons. Firstly, there are more targets near the galactic plane.
The majority of them are galactic sources but we do not know beforehand which 
ones are galactic and which are extra-galactic objects.
Secondly, the VEPS search mode observations have a shorter visible time for those
low declination sources, so there are fewer chances to pick them up in the schedules.
In order to increase the detection rate,
we plan to observe that zone in a few 8-hr segments with {\sc tianma65},
in addition to the regular CVN stations.

\begin{figure*}[htb!]
   \includegraphics[width=0.995\textwidth]{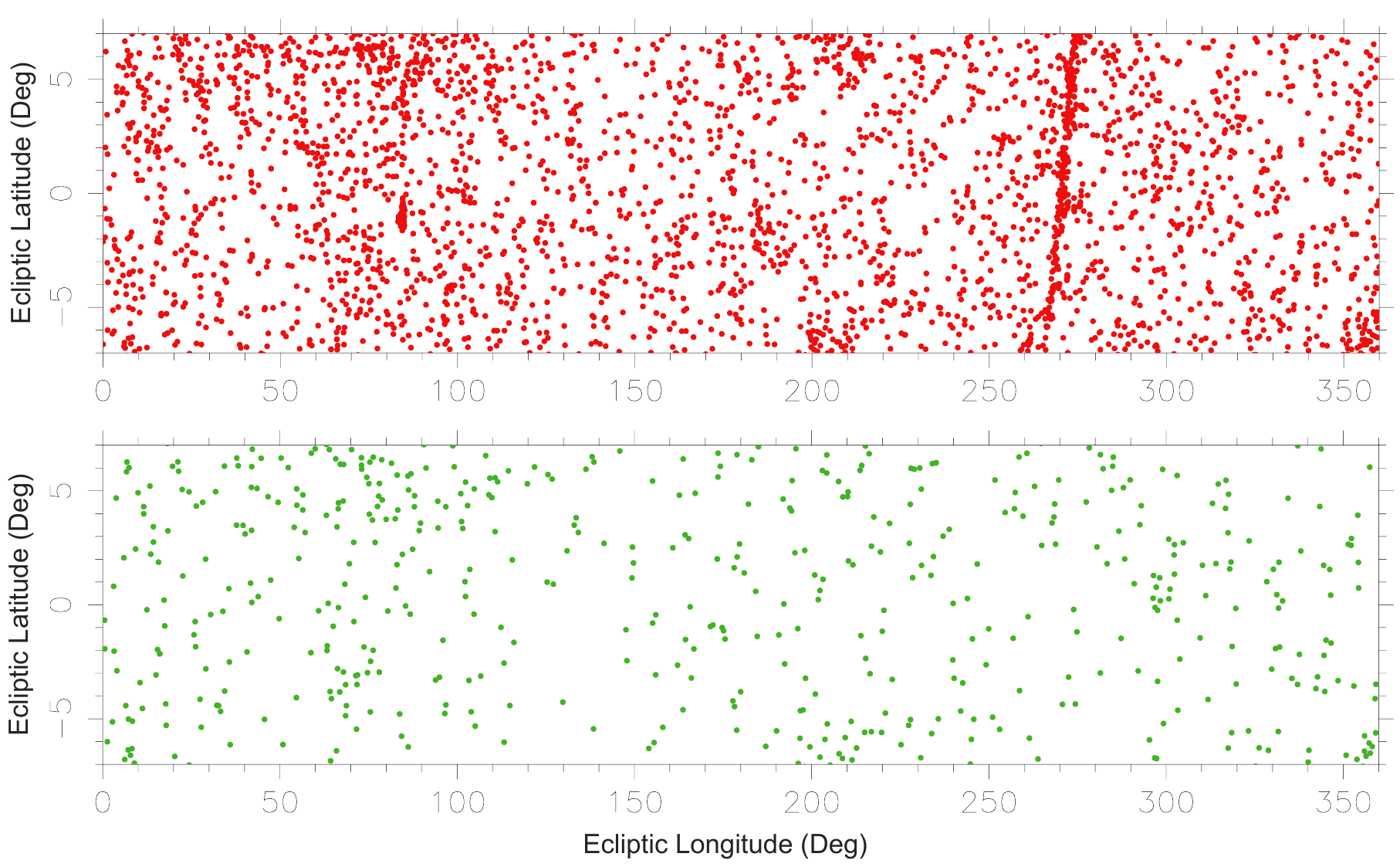}
  \caption{Distribution of ecliptic plane sources observed in the VEPS search mode.
           Top: Target sources that were observed, but not detected are shown with red circles
           (2765 objects). Bottom: Target sources that were detected are shown with green
           circles (555 objects).}
  \label{f:veps1_distr}
\end{figure*}

  Statistics of VLBI detected sources within $\pm 7.5^\circ$ of the ecliptic
plane are shown in Table~\ref{tab:ecl}. The number of known calibrators in the
ecliptic plane grew by over 60\% for two years and reached 1253 objects.
At the same time, only 29\% of the ecliptic calibrators have been determined
with position uncertainties less than 1.5~nrad using S/X or C/X dual-band VLBI.

\begin{deluxetable}{lrrr}[h!]
\tablecaption{Statistics of sources detected with VLBI within $\pm 7.5^\circ$ of the ecliptic plane.
            The second column shows the statistics before the start of the program and
            the last column shows the current numbers.
\label{tab:ecl}}
\tablecolumns{4}
\tablewidth{0pt}
\tablehead{
\colhead{} & \colhead{2015.0} & \colhead{2017.2}
}
\startdata
\# calibrators of the 1st class   &   187   &    365 \\
\# calibrators of the 2nd class   &   729   &    994 \\
\# calibrators of the 3rd class   &   768   &   1253 \\
\# non-calibrators                &   386   &    757 \\
Total \# all sources              &  1154   &   2010 \\
\enddata
\tablecomments{
A calibrator of the 1st class, the 2nd class and the 3rd class
has flux density greater than 30~mJy and a position uncertainty
less than 1.5 nrad, 10 nrad, and 100 nrad respectively.
}
\end{deluxetable}

  Table \ref{t:acc} shows the statistics of the semimajor error ellipse axes
of the position estimates of the sources that were observed in four modes.
The position accuracy derived from the BS250 is 65\% below the goal.
Approximately a factor of two increase in on-source time is needed
to reach the 1.5~nrad accuracy goal. But we should note that the target
sources for the BS250 campaign were the weakest among the ecliptic plane calibrators.

\begin{table}[h]
   \centering
   \caption{Position accuracy of observations in four different modes: VEPS Search
            mode, IVS refinement, VLBA Calibrator Survey VCS--9, and dedicated VLBA
            astrometry experiment BS250. The third and fourth columns show the 50th
            and 80th percentiles for the semimajor error ellipse in nrad respectively.}
   \par\medskip\par
   \begin{tabular}{lrrr}
      \hline
      \hline
      Mode   & \# Src & 50p & 80p \\
      \hline
      Search &    555 & 20.7 & 38.9 \\
      IVS    &     32 &  5.3 &  9.3 \\
      VCS--9 &    109 &  5.2 &  8.1 \\
      BS250  &    108 &  1.9 &  2.5 \\
      \hline
   \end{tabular}
   \label{t:acc}
\end{table}

  The accuracy of the source positions derived from IVS experiments AOV010
and AUA012 was worse than that of the VLBA experiments. On one hand,
the VLBA has ten identical sensitive antennas which can
form 45 baselines. As a comparison, a ten-station IVS network is inhomogeneous,
with one or two big antennas and more small antennas. Such a configuration
significantly reduces the number of baselines which can observe weak target
sources. On the other hand, the VLBA using a 2~Gbps recording data rate could
bring better results. Another reason is that the geodetic scheduling strategy
is not very suitable for astrometry projects. The scheduling software {\sc sked}
has a tendency to split observations into subarrays, which is acceptable for
geodesy when very strong sources are observed, but detrimental for the astrometry
of weak targets. The software also tends to schedule two consecutive scans
with a large antenna slew angle, which reduces the observation efficiency and
results in fewer scans in one session. Hence, the scheduling strategy needs to be
improved in future astrometric observations.

\section{Concluding remarks}

  The VEPS program is underway to search for all suitable calibrator
sources in the ecliptic plane with a 3--4 element VLBI network. Over 3000 target sources
have been observed in the search mode in 13 sessions, for a total of 310 hr.
We have detected 555 ecliptic plane sources with VLBI for the first time.
The detection limit in the search mode was below 20~mJy, which is sufficient
for the goal of phase referencing observations. These results demonstrate
the validity of our approach. We have reobserved 249 sources with the VLBA
and the IVS network, and improved their position estimates. However,
these observations were not sufficient to reach the goal of 1.5~nrad
position accuracy: only 29\% of the ecliptic plane calibrators have position accuracies
at that level.

In the next step, we plan to continue the observations of remaining target sources
in the search mode, and improve positions of ecliptic calibrators
to the 1.5~nrad level.
We estimate that approximately 250 hr of observing time at the
VLBA or IVS is needed to obtain the positions of all 1253 ecliptic plane calibrators
to that level, and around 150 hr more for reobservations of anticipated
400 calibrators that will be found upon completion of the VEPS program in
the search mode.

\acknowledgments

This project is supported by the National Natural Science Foundation of
China (U1331205, 11573056). This work made use of the Swinburne University of Technology
software correlator, DiFX, developed as part of the Australian Major National
Research Facilities Programme and operated under license. The Chinese VLBI Network
is operated by the Shanghai Astronomical Observatory, Chinese Academy of Sciences.
The National Radio Astronomy Observatory is a facility of the National Science Foundation operated
under cooperative agreement by Associated Universities, Inc.

\facilities{CVN, AOV, VLBA, Parkes, TMRT}



\end{document}